\documentclass[aps,prb,preprint,showkeys,superscriptaddress]{revtex4-2}

\usepackage{graphicx}
\usepackage{amsmath}
\usepackage{wrapfig}
\usepackage{multirow}

\usepackage{color,soul}
\usepackage{amssymb}
\usepackage{textcomp}
\usepackage{gensymb}
\usepackage{booktabs}
\usepackage{tablefootnote}

\usepackage[version=3]{mhchem} 

\begin{document}

\title{A comparison of the spin-phonon behaviour of \ce{Fe2P}-based magnetocaloric materials}

\author{Mikael S. Andersson}
\affiliation{Department of Chemistry - \AA{}ngstr\"{o}m Laboratory, Uppsala University, Box 538, 751 21 Uppsala, Sweden.}
\author{Simon R. Larsen}
\affiliation{Department of Chemistry - \AA{}ngstr\"{o}m Laboratory, Uppsala University, Box 538, 751 21 Uppsala, Sweden.}
\affiliation{National Institute for Materials Science (NIMS), Tsukuba, Japan.}
\author{Erna K. Delczeg-Czirjak}
\affiliation{Division of Materials Theory, Department of Physics and Astronomy, Uppsala University, Box 516, SE-751 20 Uppsala, Sweden}
\affiliation{WISE - Wallenberg Initiative Materials Science for Sustainability, Department of Physics and Astronomy, Uppsala University, SE-751 20 Uppsala, Sweden}
\author{Antonio Corona}
\affiliation{Department of Materials Science and Engineering, Uppsala University, Box 35, 751 03 Uppsala, Sweden}
\author{Jacques Ollivier}
\affiliation{Institut Laue-Langevin, BP 156, 38042 Grenoble Cedex 9, France.}
\author{Wiebke Lohstroh}
\affiliation{Technische Universit\"at M\"unchen, Garching bei M\"unchen, Heinz Maier-Leibnitz Zentrum (MLZ), Lichtenbergstr. 185748 Garching, Germany}
\author{Helen Y. Playford}
\affiliation{ISIS Pulsed Neutron \& Muon Facility, Rutherford Appleton Laboratory, Harwell Campus, OX11 0QX, United Kingdom.}
\author{Cheng Li}
\affiliation{Neutron Scattering Division, Oak Ridge National Laboratory, 1 Bethel Valley Road, Oak Ridge, TN 37830, USA.}
\author{Pascale P. Deen}
\email[Corresponding author ]{pascale.deen@ess.eu}
\affiliation{European Spallation Source ESS ERIC, Box 176, 221 00 Lund, Sweden.}
\affiliation{Nanoscience Center, Niels Bohr Institute, University of Copenhagen, 2100 Copenhagen {\O}, Denmark.}
\author{Johan Cedervall}
\email[Corresponding author ]{johan.cedervall@kemi.uu.se}
\affiliation{Department of Chemistry - \AA{}ngstr\"{o}m Laboratory, Uppsala University, Box 538, 751 21 Uppsala, Sweden.}


\begin{abstract}
Magnetic refrigeration can provide an environmentally friendly technology to reduce significantly the energy consumption of cooling devices without using harmful gases. To retain the sustainability of the device, all parts must be made from abundant materials, excluding e.g. rare earth elements. As such, materials based on \ce{Fe2P} have shown great potential for magnetocaloric devices, particularly enhanced by the pliant tunability arising through chemical substitutions. In this study, two magnetocaloric compounds, \ce{Fe2P} and \ce{FeMnP_{0.55}Si_{0.45}}, have been studied using magnetometry, neutron diffraction, theoretical modelling and inelastic neutron scattering with the aim to understand the ferromagnetic transition, closely related to the magnetocaloric effect. Analysis of the diffraction data of \ce{Fe2P} showed that it is the Fe$_{3g}$-site that drives the magnetic transition as the Fe$_{3f}$ does not have any magnetic contribution at the magnetic transition temperature. For \ce{FeMnP_{0.55}Si_{0.45}}, the magnetic transition is more gradual with coexistence of the para- and ferromagnetic phases close to the magnetic transition. In the ferromagnetic phase of \ce{FeMnP_{0.55}Si_{0.45}}, magnetic moments are observed for both sites. The temperature dependent magnetic structure behaviour found from neutron diffraction are well in agreement with our first principles calculations.

Both \ce{Fe2P} and \ce{FeMnP_{0.55}Si_{0.45}} showed two distinct regions, at different length scales, in their S(\textbf{Q},$\omega$) spectra. The two length scales can be modelled using a different set of magnetic spin states (S), using S$\rm _{Fe}$~=~2 and S$\rm _{Mn}$~=~2.5, consistent with the ground state of the magnetic atoms in \ce{(Fe{,}Mn)2(P{,}Si)}-compounds. QENS at low Q (Q~\textless{}~0.5~\AA{}) shows similar magnetic processes in both compounds with uncorrelated magnetism below the magnetic transition temperature. The occurrence of the uncorrelated state in both \ce{Fe2P} and \ce{FeMnP_{0.55}Si_{0.45}} highlights that the magnetic anisotropy does not play a major role in the formation of this state. Furthermore, this emphasises the existence of a two part system in \ce{FeMn(P{,}Si)}-based compounds, that drives the magnetic transition and in turn the magnetocaloric effect. This means that short range magnetic clusters exist above and below the magnetic ordering temperature of the compound, strongly influencing the magnetic behaviour.
\end{abstract}

\maketitle

\section{Introduction}
Magnetic refrigeration has been envisaged as an effective cooling phenomenon since the discovery of the giant magnetocaloric effect in \ce{Gd5Si2Ge2} \cite{Pecharsky1997}. Several materials have been investigated with respect to their refrigeration properties \cite{Gschneidner2000,Bruck2005,GschneidnerJr2008}, and prototypes with the most prominent class of materials have been made \cite{Franco2018} often involving rare earth metals which, in addition to being scarce, pose significant environmental and health risks. These risks include toxic waste generation, potential for radioactive contamination, and harmful effects on human health. As such, it is crucial that magnetocaloric devices are developed using resources that are plentiful and not harmful \cite{Gauss2017}. Therefore, only compounds with suitable elements from an environmental and sustainability point of view should be considered. A class of material that meets these sustainability criteria is \ce{Fe_{2-x}Mn_xP_{1-y}Si_y} with \ce{Fe2P} as the basic compound with a ferromagnetic transition temperature at 220 K, that can be tuned to a room temperature transition temperature, therefore creating a more energy efficient material, through the introduction of Mn and Si. 

\begin{figure}[t!bp]
	\centering
        \includegraphics[width=0.99\textwidth]{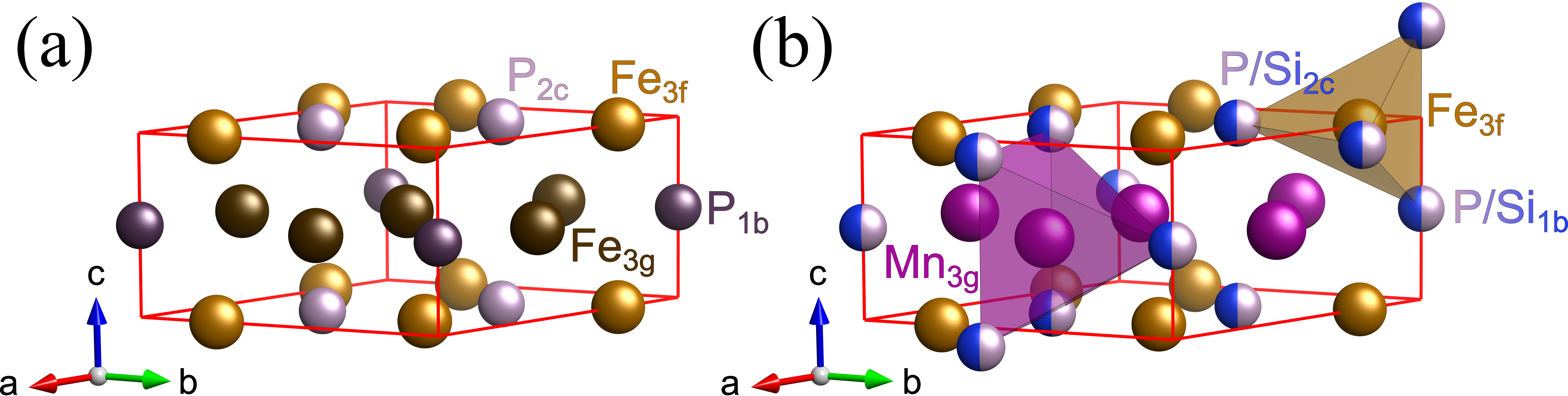}
	\caption{The nuclear structure for a) \ce{Fe2P} and b) \ce{FeMnP_{0.5}Si_{0.5}}, where the colours brown, purple, pale pink and blue corresponds to Fe, Mn, P and Si, respectively. Light brown and dark brown(purple) highlights the Fe$_{3f}$ and Fe(Mn)$_{3g}$ positions, respectively.}
	\label{fig:nuclear}
\end{figure}

\ce{Fe2P} orders in a hexagonal structure ($P\bar{6}2m$) with two crystallographic atomic positions for Fe (Wyckoff 3$f$ and 3$g$) and two for P (Wyckoff 1$b$ and 2$c$), respectively \cite{Rundqvist1959}, visualised in Figure~\ref{fig:nuclear}~a) where Fe is light and dark brown and P is pink/dark purple. Upon cooling below 220~K, \ce{Fe2P} undergoes a first-order ferromagnetic transition \cite{Lundgren1978} where the magnetic moments on the Fe sites align along the crystallographic $c$-axis \cite{Scheerlinck1978}. Given that the Curie temperature (T$\rm _C$) for \ce{Fe2P} is significantly below 300 K, the usefulness in a room temperature application is limited. However, atomic substitutions can easily be made in \ce{Fe2P} and improve the relevant properties. Upon substitutions to \ce{Fe_{2-x}Mn_xP_{1-y}Si_y}, the unit cell parameters are affected varying the strength of the magnetic moments and, importantly, enabling T$\rm _C$ to be tuned \cite{Hoglin2011,Caron2013,Miao2014,Hoglin2015,Miao2016}. 
In particular, T$\rm _C$ can be tuned from  220 K in \ce{Fe2P} to 400 K in \ce{FeMnP_{0.5}Si_{0.5}} and the ferromagnetic spin configuration in \ce{FeMnP_{0.5}Si_{0.5}} changes to the $ab$-plane~\cite{Hoglin2011}.

In the substituted compounds, strong structural preference is observed for the arrangement of Fe and Mn, with Mn preferring to occupy the pyramidal $3g$ site, leaving Fe in the tetrahedral $3f$ site, referred to as Fe(Mn)$_{3g}$ and Fe$_{3f}$, respectively \cite{Miao2014,Hoglin2011}. P and Si occupy the $1b$ and $2c$ sites with no experimental evidence for preferential occupation. The arrangement is shown in Figure~\ref{fig:nuclear}~b), which highlights the tetrahedral arrangement for Fe and the pyramidal surrounding from Mn.

A magnetic phase transition will induce a large entropic change and this is particularly pronounced for a paramagnetic to ferromagnetic transition at T$\rm _C$. The entropy (S) of a material is strongly affected by changes in internal degrees of freedom that include phonon modes, spin waves and phonon-electron interactions. As such, understanding the interplay between spin, electron, and lattice degrees of freedom is essential to optimise magnetocaloric devices. 

Intrinsic spin-phonon coupling has previously been studied for magnetocaloric materials \ce{Mn5Si3}, \ce{MnFe4Si3}, \ce{Fe2P} and \ce{LaFe_{13-x}Si_x} using inelastic neutron scattering (INS) \cite{Biniskos2017,Biniskos2018,Cedervall2019,Zhang2021}. Common findings from these studies suggest that anisotropy plays a big role in the the magnetocaloric response regardless of the nature of the magnetic ordering.  Our previous work on \ce{Fe2P} revealed that in addition to  long range magnetic correlations, and concomitant spin waves, a weakly antiferromagnetic correlated nanoscale clustered state exists below T$\rm _C$ that develops into ferromagnetic correlated clusters above T$\rm _C$. These clustered states appear to drive, most unusually, a gapped phonon state that may enhance electron mobility which in turn will enhance the magnetocaloric effect \cite{Cedervall2019}.  In addition, the long range magnetic spin structure shows a canting at T$\rm _C$. The previous INS studies all suggest that more research in this area are needed to understand the complex interactions responsible for the magnetocaloric effect.

In this study, we aim to explore the magnetic transition in materials used in magnetic refrigeration to gain insight into the magnetocaloric behaviour. This is performed by comparing the properties of \ce{Fe2P} and \ce{FeMnP_{0.55}Si_{0.45}} using mainly neutron diffraction and inelastic neutron scattering. Neutron scattering possess a unique probe that enables the exploration of the spatial and energy scales of the magnetic spin correlations and the magnon-phonon properties close to the magnetic transitions. We study the effect of substitution on the ferromagnetic transition and relevant magnetic exchange interactions that drives the magnetocaloric effect.

\section{Methods}
\subsection{Synthesis and sample quality}
Powder samples were produced using the drop synthesis method, where the volatile elements (P, Mn) are dropped into a melt of the non-volatile elements (Fe, Si) \cite{Carlsson1973}. All elements used had purities $>$99.99\%. The samples were heat treated in evacuated silica ampoules at 1273~K for 7-14 days, yielding polycrystalline samples. Crystalline quality was determined using X-ray diffraction and reveal high purity samples with no detected secondary phases. 

\subsection{Magnetometry}
Magnetic characterisation of the samples was carried out using a Quantum Design MPMS SQUID magnetometer and a Quantum Design MPMS3 SQUID magnetometer. The temperature dependence of the magnetisation has been probed using an applied magnetic field of H~=~4~kA/m ($\mu_0H$~=~5~mT). The field dependence was studied in the field range of $\pm$4000~kA/m ($\mu_0H$~=~$\pm$5~T) at select temperatures.

\subsection{Neutron diffraction}
Neutron powder diffraction (NPD) experiments were carried out at the instruments Polaris \cite{Polaris,ISISdata} at ISIS Neutron and Muon Source (Didcot, UK) and Nomad \cite{Nomad} at the Spallation Neutron Source (Oak Ridge, USA). The experiments were carried out at a temperature range from 150 to 450~K (T/T$\rm _C$ range of 0.4 to 1.2). The received diffraction patterns were analysed using the Rietveld method \cite{Rietveld1969} implemented in the software FullProf \cite{Rodriguez-Carvajal1993}. All possible magnetic space groups were evaluated with k-SUBGROUPSMAG \cite{Perez-Mato2015}, and tested to the diffraction data below T$\rm _C$.

\subsection{Inelastic neutron scattering}
Inelastic  neutron scattering  studies (INS) were  performed using  the cold  chopper  spectrometers at  Institute  Laue  Langevin and Heinz Maier-Leibnitz Zentrum (MLZ), IN5 \cite{ExpdataILL,IN5data} and TOFTOF \cite{TOFTOF}. The double partial differential scattering cross section, S(Q, $\omega$) was measured for \ce{Fe2P} and \ce{FeMnP_{0.55}Si_{0.45}} across a temperature range 0.4~$\leq$~T/T$\rm _C$~$\leq$~1.2, using neutron wavelengths of 5~\AA{} (TOFTOF) and 4.8~\AA{} (IN5), which both have instrumental energy resolutions of $\sim$90 $\mu$eV determined from the incoherent scattering of a vanadium reference.  The low Q region (0.3 to 0.5 \AA$^{-1}$) of the measured INS spectra was fitted to the incoherent scattering function

\begin{equation}
\begin{split}
S(Q,\omega)_{} =  R(Q,\omega) \otimes [\delta(\omega)A_\text{E}(Q) \\+ \sum{L_i(\omega) A_{\text{QE},i}(Q)}] + \text{Bkg}(Q),
\label{eq:Scattering function}
\end{split}
\end{equation}
where $E$ = $\hbar\omega$ is the neutron energy transfer, $\hbar$ is the Planck constant/($2\pi$), $\omega$ is the angular frequency, $\delta$ is a delta function, $
L_i$'s are Lorentzian functions used to describe the quasielastic scattering, $A_\text{E}$ and $A_{\text{QE}, i}$ are the integrated intensity corresponding to the respective delta and Lorentzian functions, $R(Q,\omega)$ is the instrument resolution function, and $\text{Bkg}(Q)$ is a linear background. The fits were performed using PAN, which is part of the DAVE distribution \cite{DAVE}.

\subsection{Theoretical modelling}
\subsubsection{Magnetic exchange parameters}
\begin{figure}[tbh]
	\centering
		\includegraphics[width=0.99\textwidth]{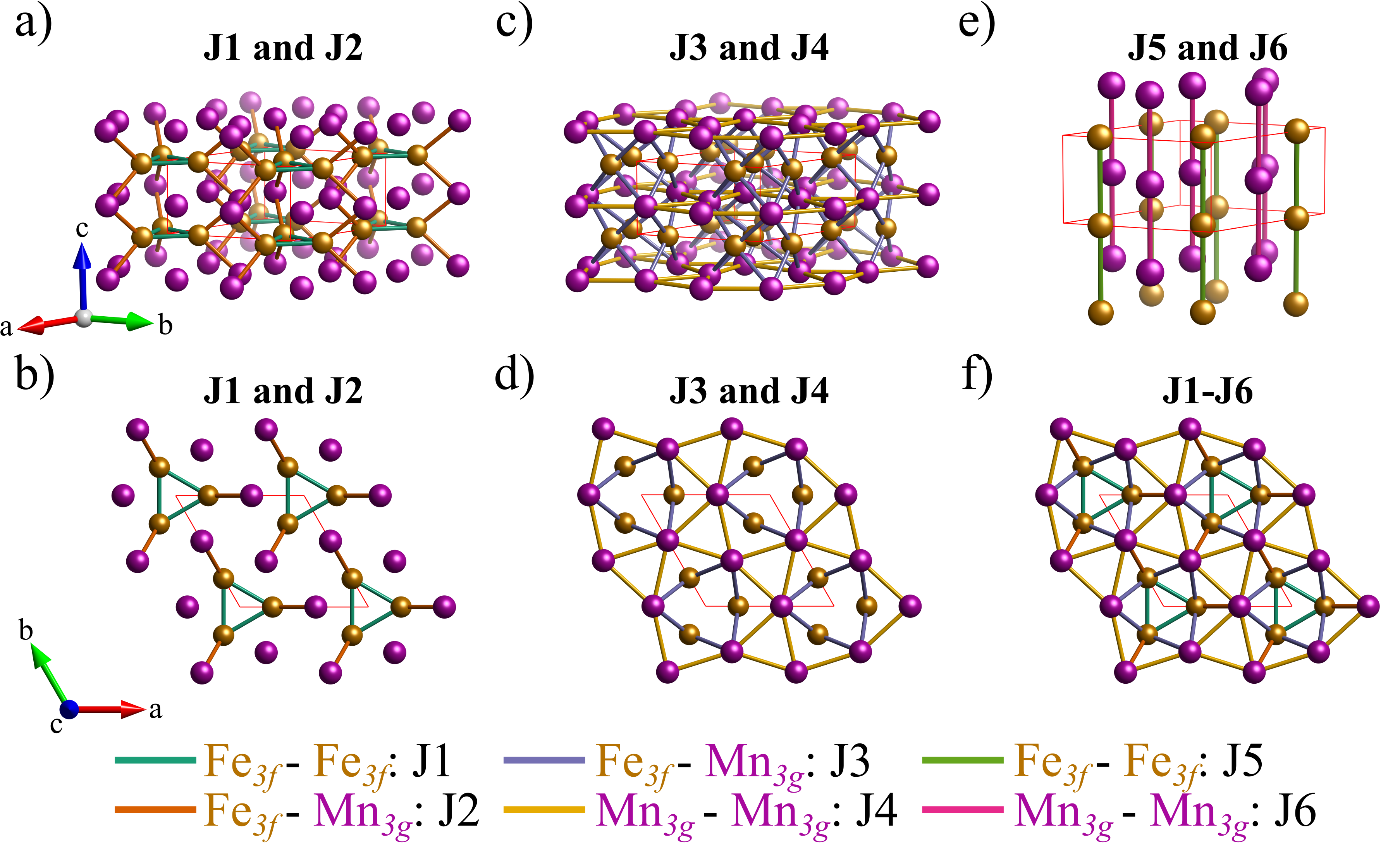}
	\caption{The magnetic interactions J1 to J6 in the \ce{Fe2P}-type structure. a) and b) shows J1 and J2 in two different projections, c) and d) shows J3 and J4 in two different projections, e) show J5 and J6 and f) show J1-J6 along $c$.}
	\label{fig:Jij}
\end{figure}

The magnetic exchange interactions were calculated for the FM experimental structures of \ce{Fe2P} and \ce{FeMnP_{0.55}Si_{0.45}} (see Table~\ref{tab:cryst}) at 0 K within the magnetic force theorem \cite{Liechtenstein1984} as it is implemented in the Lyngby version of the exact muffin-tin orbital (EMTO) code~\cite{Ruban2016}. The chemical disorder for P and Si was treated within the coherent potential approximation (CPA) \cite{Soven1967, Gyorffy1972} (EMTO-CPA \cite{Vitos2001}). The one-electron Kohn-Sham equations were solved within the soft-core and scalar-relativistic approximations, with $l_{\rm max} = 3$ for partial waves and $l_{\rm max}^{\rm t} = 5$ for their "tails". The Green's function was calculated for 16 complex energy points distributed exponentially on a semi-circular contour including states within 1 Ry below the Fermi level. The exchange-correlation effects was described within the local spin-density approximation \cite{LDA1, LDA2}. Previous theoretical studies on \ce{Fe2P}~\cite{PhysRevB.85.224435} and \ce{FeMnP_{1-x}Si_{x}} \cite{10.1063/1.4905270, 10.1063/1.4936835, Li_2016} show an oscillatory behaviour for $\mathbf{J_{i,j}}$'s with distance. It has been shown that $\mathbf{J_{i,j}}$'s calculated and summed up to distances of 2$a$ leaves the magneto-structural effects invariant~\cite{PhysRevB.85.224435} for \ce{Fe2P} and \ce{Fe2P_{1-x}Si_{x}}. J's up to the sixth nearest neighbour, J$_{6}$, are considered with higher terms not further affecting the results (Figure~\ref{fig:Jij}).

\subsubsection{INS spectra from linear spin wave theory}
Linear spin wave theory was used to simulate the magnetic contributions of the INS spectra for \ce{Fe2P} and \ce{FeMnP_{0.55}Si_{0.45}} using the exchange interactions determined by the magnetic force theorem; see Table~\ref{tab:Jij} and Figure~\ref{fig:Jij}. The simulations were made using the SpinW software \cite{SpinW}, to compute the INS spectra based on the Hamiltonian  
\begin{equation}
\mathcal{H}=\sum_{i,j}\mathbf{S}_iJ_{ij}\mathbf{S}_j + \sum_i \mathbf{S}_iA_i\mathbf{S}_i ,    
\end{equation}
where $S_i$ are spin vector operators with $\mathbf{J_{i,j}}$ the coupling between the spins $S_i$ and $S_j$. Anisotropy is introduced via $\mathbf{A_{i,j}}$, a 3$\times$3 easy-axis anisotropy matrix. In the case of \ce{Fe2P} all simulated spectra used a spin state of S = 2 for both the Fe$_{3f}$ and Fe$_{3g}$ sites as suggested by previous polarized neutron diffraction experiments~\cite{Fujii1979} and a previously computed uniaxial anisotropy of D = 0.1088 meV/atom along the $c$-axis~\cite{Cedervall2019}. For \ce{FeMnP_{0.55}Si_{0.45}} two different spin configurations (S$\rm _{Fe}$~=~2 and S$\rm _{Mn}$~=~2; S${\rm _{Fe}}$~=~2 and S${\rm _{Mn}}$~=~2.5) and a uniaxial anisotropy of D = 0 meV/atom were used to simulate the INS spectra. No polarized neutron diffraction experiments for Mn substituted compounds of \ce{Fe2P} have been performed to date and thus the spin state for Mn is unknown, therefore, the spin states have to be estimated to model the INS spectrum of \ce{FeMnP_{0.55}Si_{0.45}}. The spin states are connected to the oxidation state of Fe and Mn and an estimation of the spin states can be made considering the oxidation states for Fe and Mn. Only S$\rm _{Fe}$~=~2 and S$\rm _{Mn}$~=~2; S${\rm _{Fe}}$~=~2 and S${\rm _{Mn}}$~=~2.5 give reasonable oxidation states for Fe and Mn and thus only these two spin state configurations are considered in this study.

\begin{table*}[tbh]
\small
\caption{Atomic distances and J-values for \ce{Fe2P} and \ce{FeMnP_{0.55}Si_{0.45}} used as input for linear spin wave theory simulations.}
\begin{tabular*}{0.85\textwidth}{@{\extracolsep{\fill}}l|c|c|c|c|c}
\multicolumn{2}{c|}{} & \multicolumn{2}{c|}{\ce{Fe2P}} & \multicolumn{2}{c}{\ce{FeMnP_{0.55}Si_{0.45}}} \\
J & Atom pair & Distance (\AA) & J-value (meV) & Distance (\AA) & J-value (meV) \\
\hline
1 & Fe$_{3f}$-Fe$_{3f}$ & 2.6070 & -4.40 & 2.7380 & -1.35 \\
2 & Fe$_{3f}$-Fe(Mn)$_{3g}$ & 2.6150 & -11.20 & 2.6630 & -20.77 \\
3 & Fe$_{3f}$-Fe(Mn)$_{3g}$ & 2.7230 & -8.30 & 2.7650 & -17.53 \\
4 & Fe(Mn)$_{3g}$-Fe(Mn)$_{3g}$ & 3.0760 & -14.70 & 3.2630 & -9.74 \\
5 & Fe$_{3f}$-Fe$_{3f}$ & 3.4590 & -1.30 & 3.2810 & -9.95 \\
6 & Fe(Mn)$_{3g}$-Fe(Mn)$_{3g}$ & 3.4590 & -2.45 & 3.2810 & -10.87 \\
\end{tabular*}
\label{tab:Jij}
\end{table*}

\section{Results and discussion}
\subsection{Magnetometry}
\begin{figure}[tbh]
	\centering
		\includegraphics[width=0.99\textwidth]{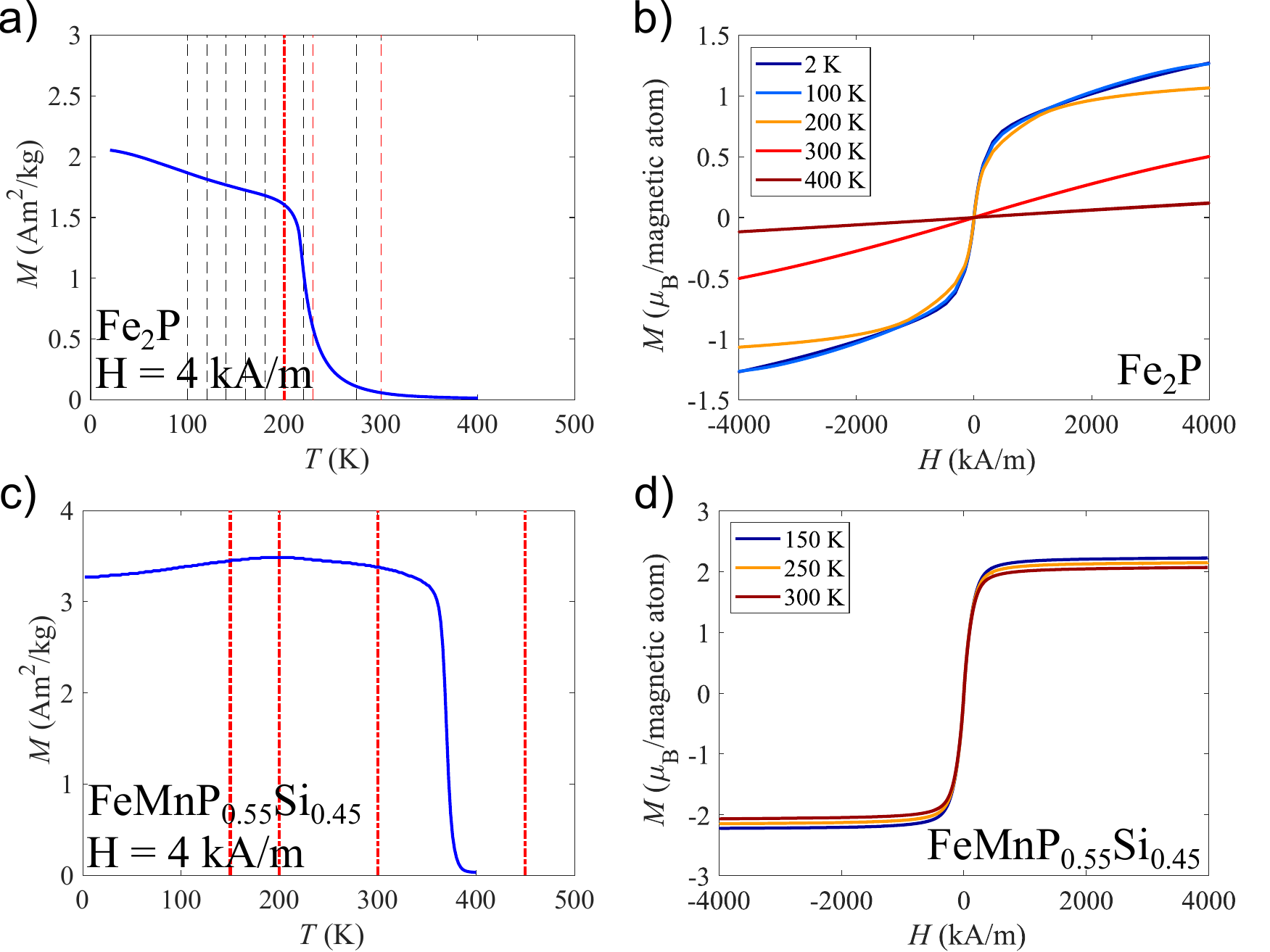}
	\caption{Magnetisation as a function of temperature for a) \ce{Fe2P} and c) \ce{FeMnP_{0.55}Si_{0.45}}. The dashed black lines indicate INS measurements temperatures, the dashed red lines NPD measurement temperatures. The thicker red dash-dotted lines indicates both INS and NPD measurements. Magnetisation as a function of applied magnetic field for b) \ce{Fe2P} and d) \ce{FeMnP_{0.55}Si_{0.45}}}
	\label{fig:Magnetometry}
\end{figure}

Magnetisation as a function of temperature for \ce{Fe2P} and \ce{FeMnP_{0.55}Si_{0.45}} are presented if Figure~\ref{fig:Magnetometry} a) and b).  As can be observed \ce{Fe2P} exhibits a magnetic phase transition at $\sim$220 K, while \ce{FeMnP_{0.55}Si_{0.45}} exhibits a transition at $\sim$370 K, which is in good agreement with previous studies~\cite{Lundgren1978,Hoglin2015}. The field dependence of the magnetisation for \ce{Fe2P} and \ce{FeMnP_{0.55}Si_{0.45}} are presented in Fig. \ref{fig:Magnetometry} c) and d). For \ce{Fe2P} it can be seen that the sample has an S shaped curve at the lowest temperatures in good agreement with previous studies which found that \ce{Fe2P} is ferromagnetic at these temperatures. However, it is clear that the sample does not fully saturate at the lowest temperatures even at 4000 kA/m ($\mu_0H$~=~5~T). This is likely an effect of the strong magnetic anisotropy ($2.32\times10^6$ J/m$^3$) in \ce{Fe2P}~\cite{fujii1977}, which prevents the magnetic moments of the individual crystallites in the powder sample from fully aligning with the field~\cite{Chikazumi2010}. At 300 K and 400 K, the curve appears to be linear indicating that the sample is in its paramagnetic phase. For \ce{FeMnP_{0.55}Si_{0.45}} the curve is more square like and exhibits a clear plateau at larger field suggesting a smaller magnetic anisotropy than in \ce{Fe2P}. From the field dependent data it is also evident that \ce{FeMnP_{0.55}Si_{0.45}} has a significantly larger average magnetic moment as compared to \ce{Fe2P}, e.g. at 4000 kA/m ($\mu_0H$~=~5~T) and T/T$\rm _C$$\sim$0.4 (150 K and 100 K respectively) the average moment is 2.25 $\mu_B$/magnetic atom for \ce{FeMnP_{0.55}Si_{0.45}} and 1.25 $\mu_B$/magnetic atom for \ce{Fe2P}.

\subsection{Nuclear and magnetic structures}
The magnetic phase transitions, magnetic moment size and corresponding structural variations across the paramagnetic (PM) to ferromagnetic (FM) phase has been probed by neutron diffraction. Figure~\ref{fig:NPD-PM} shows the neutron powder diffraction patterns, integrated over all energies, for \ce{Fe2P} (a) and \ce{FeMnP_{0.55}Si_{0.45}} (b) in the paramagnetic state at 305 and 470~K, T/T$\rm _C$~$\approx$~1.3, respectively. The ferromagnetic state was examined for \ce{Fe2P} and \ce{FeMnP_{0.55}Si_{0.45}} at 200 (T/T$\rm _C$~$\approx$~0.9) and 150 K (T/T$\rm _C$~$\approx$~0.4), respectively. 

\subsubsection{Paramagnetic state}
\begin{figure}[tbh]
	\centering
		\includegraphics[width=0.99\textwidth]{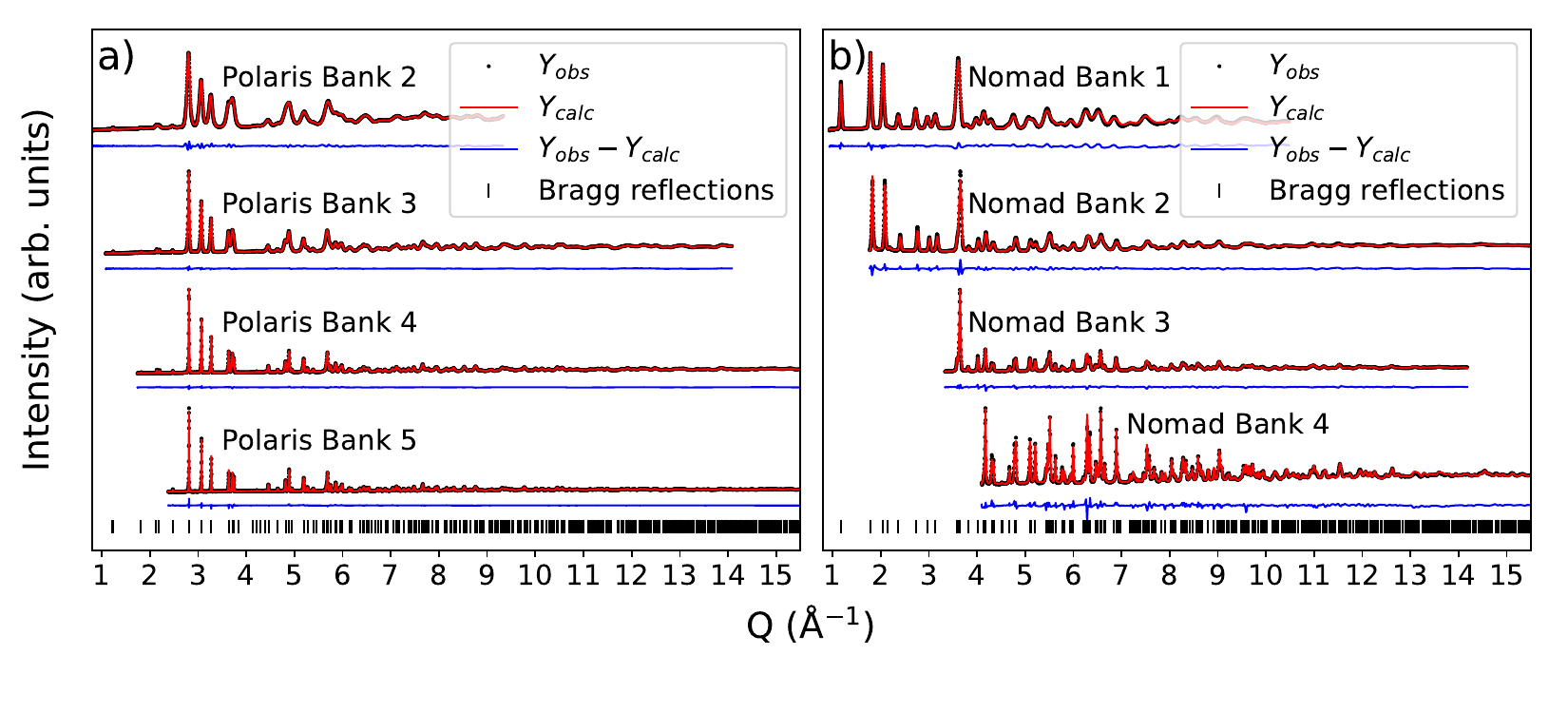}\vspace{-1cm}
	\caption{Neutron diffraction patterns for a) \ce{Fe2P} using Polaris@ISIS at 300~K ($\chi^2$~=~9.8, $R_{wp}$(Bank~5)~=~2.4) and b) \ce{FeMnP_{0.55}Si_{0.45}} using Nomad@SNS at 470~K ($\chi^2$~=~1.0, $R_{wp}$(Bank~4)~=~6.1).}
	\label{fig:NPD-PM}
\end{figure}

The neutron diffraction patterns for T $>$ T$\rm _C$ for both samples can be fully described using one \ce{Fe2P} phase ($P\bar{6}2m$), Figure~\ref{fig:NPD-PM}. For \ce{Fe2P} the refined structure perfectly corresponds to the previous sample by us~\cite{Cedervall2019} and others~\cite{Rundqvist1959,Lundgren1978,Carlsson1973}, both regarding the atomic positions and the unit cell parameters.

For \ce{FeMnP_{0.55}Si_{0.45}}, Fe almost exclusively occupies the tetragonal Fe$_{3f}$-site, refined to 92\% occupancy with Mn in the remaining 8\%. In contrast Mn occupies the pyramidal Fe(Mn)$_{3g}$-site (92\%). The occupancies are in accordance to previous neutron diffraction studies of \ce{FeMnP_{0.5}Si_{0.5}}~\cite{Miao2014,Hoglin2011}. The unit cell parameters and the shortest Fe-Fe distances from the refined models are listed in Table~\ref{tab:cryst} and full descriptions of the refined models are listed in the Supporting Materials (SM), Tables~S1 and S2~\cite{SM}. 

\begin{table*}[tbh]
\small
\caption{Unit cell parameters, Fe-Fe(Mn) nearest neighbour distances and the magnetic moments for \ce{Fe2P} and \ce{FeMnP_{0.55}Si_{0.45}} in the PM and FM states. Errors are given within parenthesis.}
\begin{tabular*}{0.8\textwidth}{@{\extracolsep{\fill}}l|cc|cc}
Compound & \multicolumn{2}{c|}{\ce{Fe2P}} & \multicolumn{2}{c}{\ce{FeMnP_{0.55}Si_{0.45}}} \\
Magnetic state & PM & FM & PM & FM \\
Temperature (K) & 300 & 200 & 470 & 150 \\
\hline
$a$ (\AA{}) & 5.8655(1) & 5.8698(1) & 6.0839(5) & 6.2243(4) \\
$c$ (\AA{}) & 3.4563(1) & 3.4430(1) & 3.4748(3) & 3.2808(2) \\
$c/a$ & 0.589(1) & 0.587(1) & 0.571(4) & 0.527(3) \\
\hline
Fe$_{3f}$-Fe$_{3g}$ (\AA{}) & 2.6120(1) & 2.6137(1) & 2.6768(3) & 2.7883(4) \\
Fe$_{3f}$-Fe(Mn)$_{3g}$ (\AA{}) & 2.6271(1) & 2.6191(1) & 2.6877(6) & 2.6722(6) \\
Fe(Mn)$_{3g}$-Fe(Mn)$_{3g}$ (\AA{}) & 3.0857(1) & 3.0845(1) & 3.1898(2) & 3.2850(6) \\
\hline
Magnetic moment Fe$_{3f}$ ($\mu \rm _B$) & - & 0.4(3) & - & 1.9(4) \\
Magnetic moment Fe(Mn)$_{3g}$ ($\mu \rm _B$) & - & 1.6(1) & - & 2.2(3)
\end{tabular*}
\label{tab:cryst}
\end{table*}

\subsubsection{Magnetic state}
\begin{figure}[tbh]
	\centering
		\includegraphics[width=0.99\textwidth]{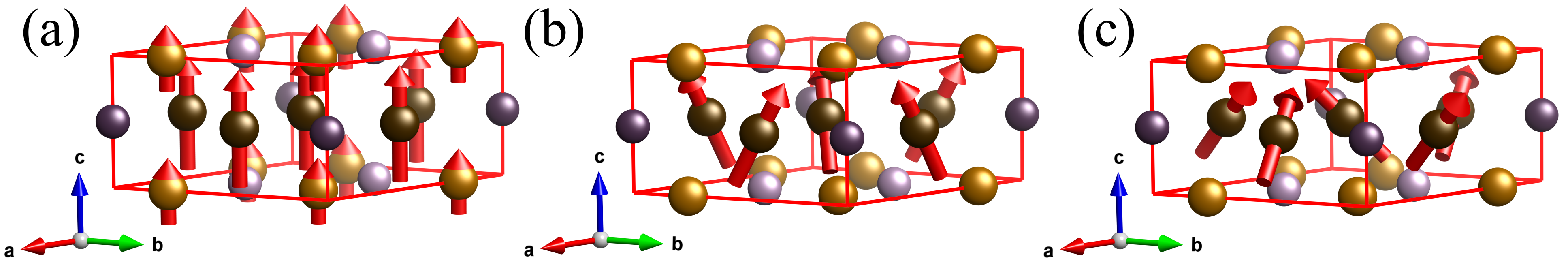}
	\caption{a) Low temperature magnetic structure for \ce{Fe2P} (MSG $P\bar{6}2'm'$). The canted magnetic structures close to T$\rm_C$ are shown for the MSGs b) $P31m'$ and c) $P32'1$ at 235~K.}
	\label{fig:magstructFe2P}
\end{figure}

At 200~K (T/T$\rm _C$~=~0.9) \ce{Fe2P} the neutron diffraction patterns can be described using a magnetic structure along the crystallographic $c$-direction (magnetic space group (MSG) $P\bar{6}2'm'$), in agreement with previous findings~\cite{Scheerlinck1978}. However, the total magnetisation at this temperature is rather low, 2.0(1)~$\mu \rm _B$/f.u., in comparison to the theoretical value of 3.1~$\mu \rm _B$/f.u. (0~K)~\cite{Delczeg2010}. The majority of the magnetic moment comes from the Fe$_{3g}$ site, refined to 1.6(1)~$\mu \rm _B$ in agreement with previous findings at liquid nitrogen temperatures~\cite{Koumina1998}. The refined values for both sites are listed in Table~\ref{tab:cryst}. As the sample is heated and approaches T$\rm _C$, the peculiar canting of the Fe moments towards the $ab$-plane \cite{Cedervall2019} is again observed here. This can be modelled using two different magnetic space groups, either $P31m'$ or $P32'1$, meaning that the symmetry is lowered from the parent space group, $P\bar{6}2m$. Common to both space groups is a large magnetic contribution along the crystallographic $c$-axis and a smaller antiferromagnetic component in the $ab$-plane. However, the current data cannot distinguish between the two magnetic models. The different magnetic structures of \ce{Fe2P} are shown in Figure~\ref{fig:magstructFe2P}. In the refinements of the magnetic moments above 200~K, the moment of Fe$_{3f}$ was unstable and yielded moments close to 0 with large errors, and was therefore omitted from the refinements. However, theoretical calculations~\cite{Delczeg2010,Delczeg-Czirjak2012July} suggests that the magnetic moment of Fe$_{3f}$ disappears during the heating as the transition is driven by the moment of Fe$_{3g}$. The results from our NPD study supports the predictions from theory and clearly shows that Fe$_{3g}$-site drives the magnetisation in \ce{Fe2P}.

\begin{figure}[tbh]
	\centering
		\includegraphics[width=0.33\textwidth]{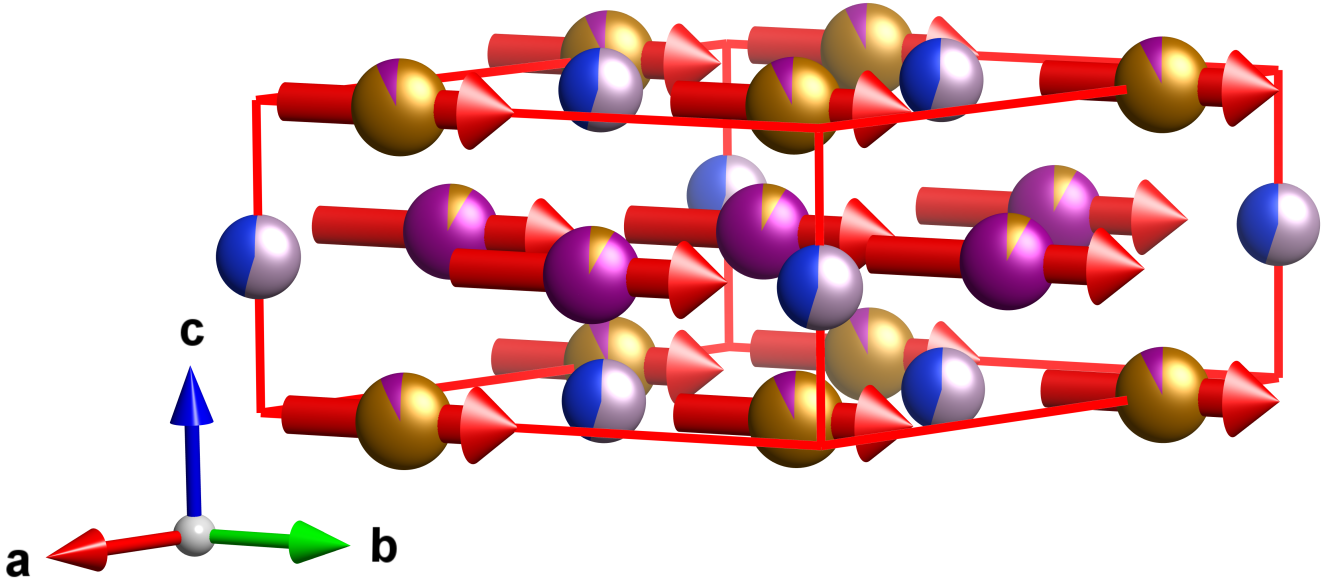}
	\caption{Low temperature magnetic structure for \ce{FeMnP_{0.55}Si_{0.45}} (MSG $Am'm'2$).}
	\label{fig:magstructFeMnPSi}
\end{figure}

The ferromagnetic structure of \ce{FeMnP_{0.55}Si_{0.45}} can be modelled using the magnetic space group $Am'm'2$, with the magnetic moments ordered along the crystallographic $b$-axis, Figure~\ref{fig:magstructFeMnPSi}. At 150~K (T/T$\rm _C$~=~0.4), the total magnetisation is refined to 4.2(1)~$\mu \rm _B$/f.u., close to the values previously reported from experiments~\cite{Hoglin2011} and theoretical calculations~\cite{Hudl2011} for \ce{FeMnP_{0.5}Si_{0.5}}. In agreement with the magnetisation measurements, the magnetic transition is found to be more gradual in \ce{FeMnP_{0.55}Si_{0.45}} compared to \ce{Fe2P}, with a coexistence of a FM and PM phase at 400~K. Furthermore, the FM phase at 400~K have magnetic moments on both Fe$_{3f}$ (1.7(6)~$\mu_B$) and Mn$_{3g}$ (1.8(5)~$\mu_B$) sites. The different temperature dependant behaviour comes from the effects of substitutions and its impact on the electronic structure. Mn substitutions gives an increased magnetic strength on both the Fe$_{3f}$ and the Mn$_{3g}$ sites, resulting in a non-zero magnetic moment on Fe$_{3f}$, as shown in SM (Figure~S1).

\subsection{Inelastic neutron scattering}
\begin{figure}[tbh]
	\centering
		\includegraphics[width=0.99\textwidth]{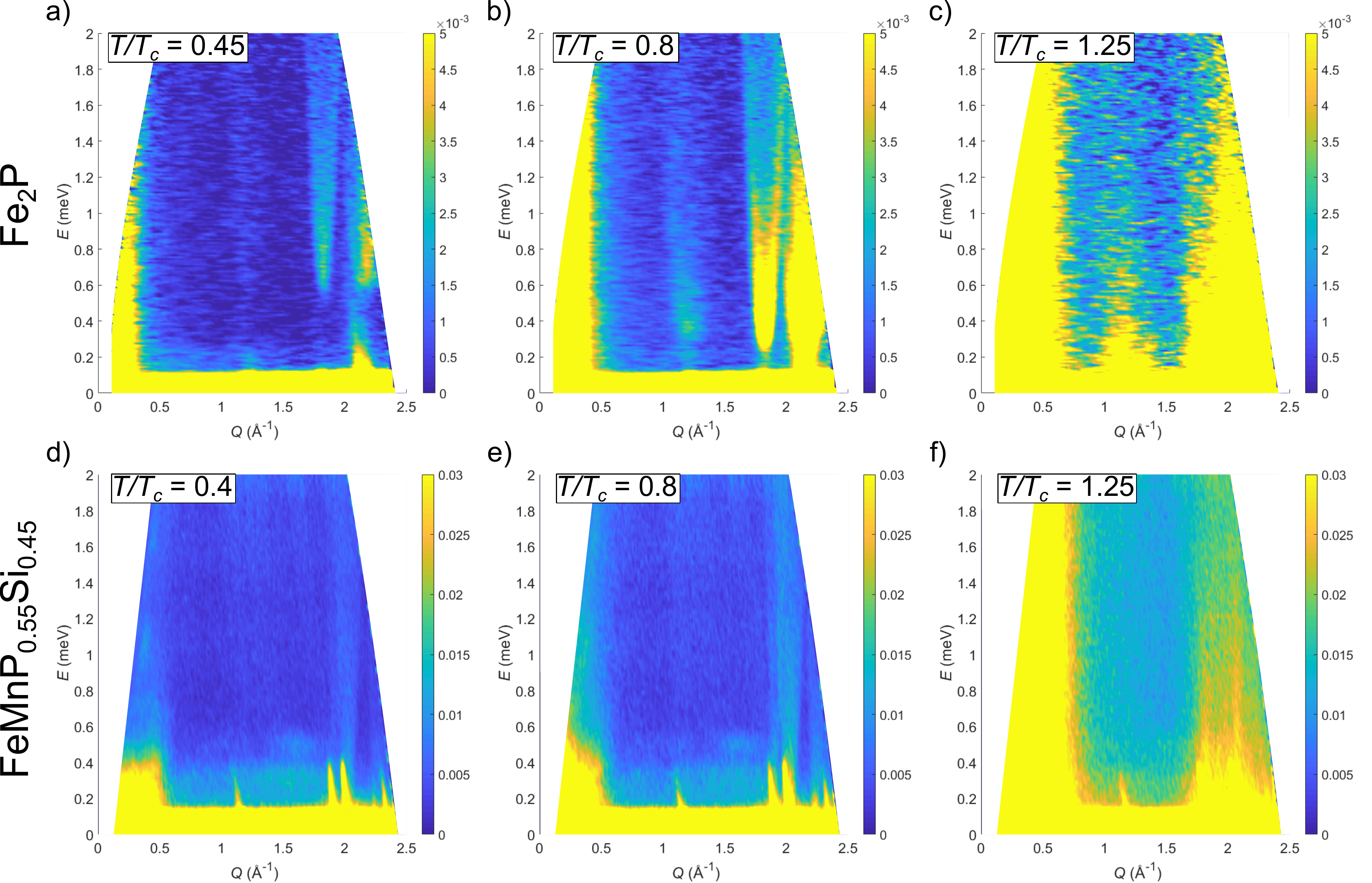}
	\caption{S(\textbf{Q},$\omega$) for a)-c) \ce{Fe2P} and d)-f) \ce{FeMnP_{0.55}Si_{0.45}} at different reduced temperatures (T/T$\rm _C$). Part of the low Q scattering for \ce{FeMnP_{0.55}Si_{0.45}} is related to scattering from the instrument. This is best observed in d).}
	\label{fig:SQW}
\end{figure}

Figure \ref{fig:SQW} shows S(\textbf{Q},$\omega$) for \ce{Fe2P} and \ce{FeMnP_{0.55}Si_{0.45}} at two temperatures below T$\rm _C$ and one above. For \ce{Fe2P} clear correlated excitations are observed at Q$\sim$1.2 \AA$^{-1}$, Q$\sim$1.8 \AA$^{-1}$ and Q$\sim$2.15 \AA$^{-1}$ corresponding to the spin-spin and phononic dynamic correlations from the (1~0~0), (0~0~1) and (1~1~1) reflections, respectively. In addition, a large broad feature at low Q (Q~\textless~0.5~\AA), and extending high in energy transfer can be observed up to 1~meV, which exists for T~$<$~T$\rm _C$ and grows significantly in intensity upon approaching - and beyond - T$\rm _C$. This feature is similar to the observed scattering behaviour in the magnetocaloric compound \ce{LaFe_{11.8}Si_{1. 2}}, and was ascribed to spin waves~\cite{morrison2024}.

Since the magnetic moments in \ce{Fe2P} are aligned along the crystallographic $c$-axis, and the neutron magnetic scattering cross section is sensitive to moments perpendicular to Q and therefore any scattering observed at (0~0~1) cannot be magnetic. As such the observed excitation at Q$\sim$1.8 \AA$^{-1}$ (0~0~1) is purely phononic \cite{Boothroyd}. The excitations at Q$\sim$1.2~\AA$^{-1}$ (1~0~0) and Q$\sim$2.15~\AA$^{-1}$ (1~1~1) contain both phonon and magnon contributions.

In the case of \ce{FeMnP_{0.55}Si_{0.45}} excitations can be observed at Q$\sim$1.8 \AA$^{-1}$ and Q$\sim$2.15 \AA$^{-1}$ and are expected to contain both phonon and magnon contributions, due to the alignment of the magnetic moment along the crystallographic $ab$-axis. Similar to \ce{Fe2P}, a purely phononic excitation is expected to be observed at Q$\sim$1.2 \AA$^{-1}$, however, no clear excitation is observed. This is not unexpected since \ce{Fe2P} has a coherent scattering cross section that is 5 times larger than \ce{FeMnP_{0.55}Si_{0.45}} and thus the phonon excitations are expected to be very weak for \ce{FeMnP_{0.55}Si_{0.45}}.  Similar to \ce{Fe2P} there is a large broad feature at low Q that grows in intensity as T$_{\rm C}$ is approached. Of note is the lack of an energy gap of the magnetic and phononic excitation for \ce{FeMnP_{0.55}Si_{0.45}}, in contrast to \ce{Fe2P} and other recent studies of \ce{(Mn{,}Fe)5Si3}~\cite{Biniskos2017,Biniskos2018} and \ce{LaFe_{13-x}Si_{x}}~\cite{Zhang2021}. A gap in the magnetic excitation spectrum is often derived from magnetic anisotropy and considered vital to drive the magnetocaloric effect (MCE). In this case the significant anisotropy in \ce{Fe2P} is strongly reduced for \ce{FeMnP_{0.55}Si_{0.45}} to below the energy resolution of the experiment, $\Delta$E~=~0.08~meV, even though the energy scales of the material have increased, from T$\rm _C$~=~220~K in \ce{Fe2P} to 370 K in \ce{FeMnP_{0.55}Si_{0.45}}. A full list of the excitation positions are found in SM (Table~S3).

\begin{figure}[tbh]
	\centering
		\includegraphics[width=0.99\textwidth]{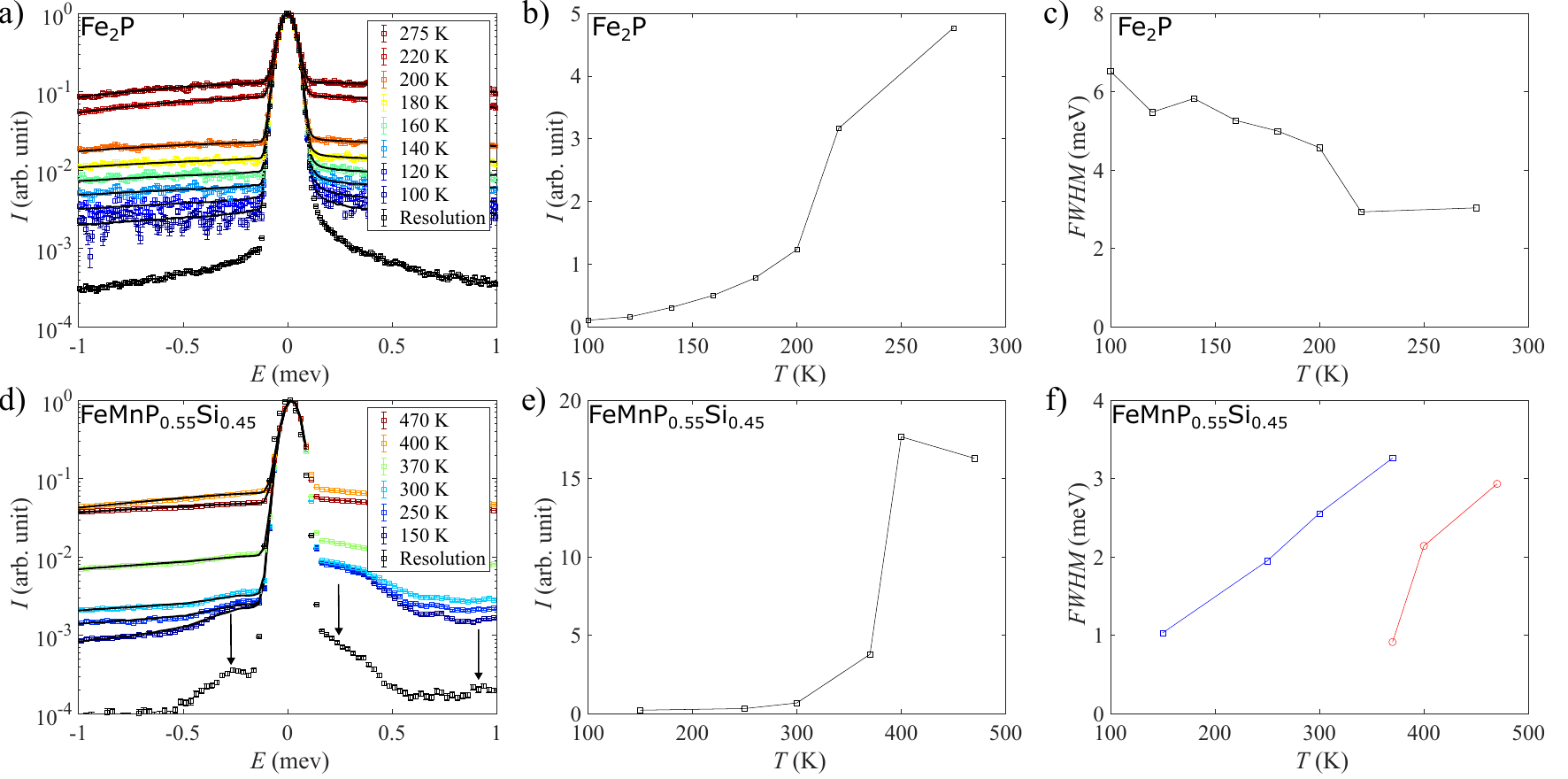}
	\caption{The temperature dependence of S(E) of the QENS-regime with Q integrated across in 0.3~\textless~Q~\textless~0.5~\AA$^{-1}$ for a) \ce{Fe2P} and d) \ce{FeMnP_{0.55}Si_{0.45}}. N.B. the spectra for \ce{FeMnP_{0.55}Si_{0.45}} contains scattering from the sample environment, which is indicated by black arrows in d).  Temperature dependence of the integrated intensity of the Lorentzian scattering for b) \ce{Fe2P} and e) \ce{FeMnP_{0.55}Si_{0.45}}. N.B. for 370 K two Lorentzian components were required to adequately describe the data for \ce{FeMnP_{0.55}Si_{0.45}}. The intensity at 370 K in e) is the sum of the intensities of both components. FWHM of the Lorentzian components for c) \ce{Fe2P} and f) \ce{FeMnP_{0.55}Si_{0.45}}. N.B. due to the occurrence of two Lorentzian components at 370 K for \ce{FeMnP_{0.55}Si_{0.45}}  both widths are shown in f).}
	\label{fig:QENS}
\end{figure}

The dynamics at low Q can be described by a Lorentzian lineshape that represents the Fourier Transform of an exponential decay function with a Full Width at Half Maximum (FWHM), $\Gamma$, that describes a characteristic decay time, $\tau$, of the magnetisation correlation function such that
\begin{equation}
\Gamma = 2\hbar/\tau .
\end{equation}
In the absence of correlated magnetic order, magnetic scattering with an exponential decay of the correlation function in time, quasi-elastic scattering, would dominate the magnetic scattering. The Fourier Transform of an exponential decay of correlations in time, and thus S($\omega$), is Lorentzian. In Figure~\ref{fig:QENS}, the low Q scattering for \ce{Fe2P} and \ce{FeMnP_{0.55}Si_{0.45}} were studied. S(\textbf{Q},$\omega$) was integrated across 0.3 $<$ Q $<$ 0.5 \AA{}$^{-1}$ and shows the normalised, to unity, scattering profile S($\omega$) compared to the instrumental resolution (Vanadium) and a fit of the data to eqn. \ref{eq:Scattering function} at each temperature. For \ce{Fe2P} the fits were made using an energy window of -1.6 to 1.3 meV, while the fits for \ce{FeMnP_{0.55}Si_{0.45}} used a window of -1.6 to 0.1 meV due to prominent features limiting fitting above 0.1 meV. As seen in Figure~\ref{fig:QENS} d) these features are also present in the Vanadium resolution, thus proving that these features are related to scattering from the sample environment and not related to the physics of \ce{FeMnP_{0.55}Si_{0.45}}.

The fits to the data comprises the instrumental resolution function, measured using Vanadium, and a Lorentzian profile. For \ce{Fe2P}, both below and above T$\rm _C$, the magnetic correlations can be accurately described by a Lorentzian function and are thus derived from uncorrelated magnetism; see Figure. \ref{fig:QENS} a). This is rather unusual for the ordered state. The  intensity of this scattering shows a pronounced jump at the transition temperature indicating that this scattering is perturbed by the long range ordered state, and possibly vice versa; see Figure \ref{fig:QENS} b). In a similar manner the FWHM of the scattering ($\tau~\propto1/\text{FWHM}$) exhibits a drastic decrease at T$\rm _C$; see Figure \ref{fig:QENS} c).

In the case of \ce{FeMnP_{0.55}Si_{0.45}}, S(\textbf{Q},$\omega$) also shows very strong low Q scattering similar to that found in \ce{Fe2P} that can be correlated to T$\rm _C$ in a similar fashion to \ce{Fe2P}; see Figure~\ref{fig:QENS}~d). In addition, due to two phases (FM and PM) being present in the transition region in  \ce{FeMnP_{0.55}Si_{0.45}} the spectra at 370 K exhibits two Lorentzian components. One of the components is attributed to the uncorrelated dynamics observed below T$\rm _C$, while the second component is attributed to the scattering of the paramagnetic phase. For both \ce{Fe2P} and \ce{FeMnP_{0.55}Si_{0.45}} it should be noted that the FWHM of the Lorentzian components are broader than or in the order of the width of the experimental energy window, which together with a relatively low intensity of the Lorentzian component at the lower temperatures introduces a large uncertainty in the width. Thus, the exact width at lower temperatures should be interpreted with caution. However, from the fitting it is clear that the width changes significantly around T$_{\rm C}$, i.e. it is not possible to fit the Lorentzian component just below T$_{\rm C}$ with a width similar to the width observed just above T$_{\rm C}$ or vice versa.

\subsubsection{Linear spin wave theory analysis}
\begin{figure}[tbh]
	\centering
		\includegraphics[width=0.99\textwidth]{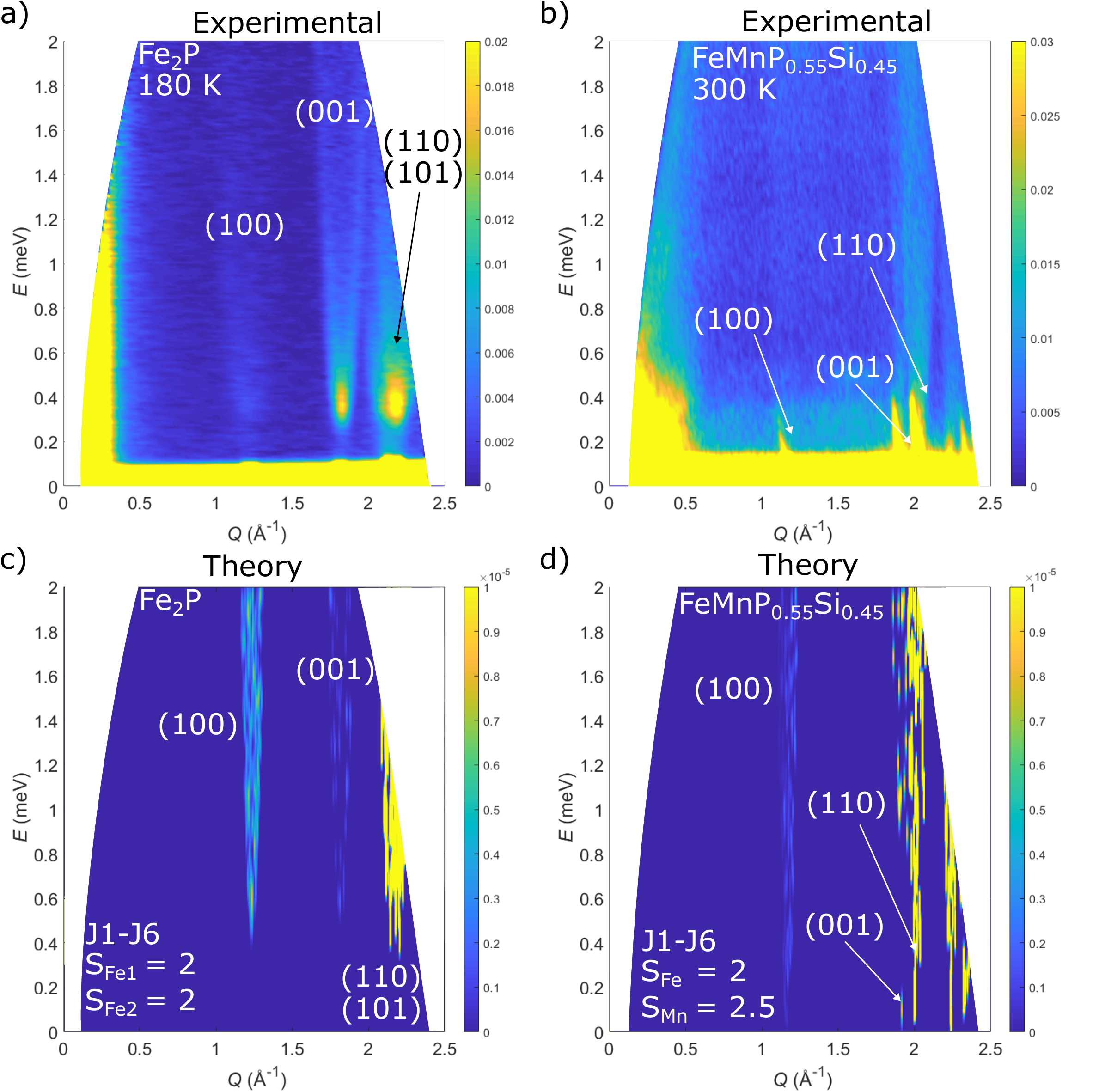}
	\caption{Experimental INS spectra for a) \ce{Fe2P} at 180 K and b) for \ce{FeMnP_{0.55}Si_{0.45}} at 300 K. Simulated INS spectra using LSWT for c) \ce{Fe2P} and for d) \ce{FeMnP_{0.55}Si_{0.45}} at 0 K.}
	\label{fig:SpinW_Fe2P}
\end{figure}

S(\textbf{Q},$\omega$) simulated spectra for \ce{Fe2P} and \ce{FeMnP_{0.55}Si_{0.45}} in the ordered regime, using the magnetic exchange parameters J1-J6, are shown in Figure \ref{fig:SpinW_Fe2P} and their corresponding experimental data at T/T$\rm _C$~$\approx$~0.8. The considered spin states, which are connected to the oxidation states, are S$\rm _{Fe}$~=~2 and S$\rm _{Mn}$~=~2 or 2.5. For \ce{Fe2P}, the simulated spectrum accurately captures both the position and the gap of the magnetic excitations in \ce{Fe2P} above $\sim$0.5~\AA$^{-1}$. It is worth noting that from the NPD analysis, the excitation at 1.8~\AA$^{-1}$ does not contain magnetic scattering and thus no strong magnetic excitation should be present in the S(\textbf{Q},$\omega$) spectra at this position. However, while the excitations above $\sim$0.5~\AA$^{-1}$ are well described by taking into account J1-J6 the low Q feature present in the experimental INS spectrum is not. The simulated INS spectra for \ce{FeMnP_{0.55}Si_{0.45}} using the two different spin configurations (S$\rm _{Fe}$~=~2 and S$\rm _{Mn}$~=~2 or 2.5) both captures the two strong excitations at $\sim$2~\AA$^{-1}$ and $\sim$2.25~\AA$^{-1}$, shown in SM (Figure~S2). The relative intensity between the two weaker excitations at $\sim$1.15~\AA$^{-1}$ and $\sim$1.9~\AA$^{-1}$ in the experimental INS spectrum is better captured by the S${\rm _{Fe}}$~=~2 and S${\rm _{Mn}}$~=~2.5 spin configuration, proving the spin states in \ce{FeMn(P{,}Si)}. The spin configurations of S$\rm _{Fe}$~=~2 and S$\rm _{Mn}$~=~2.5 are consistent with the spin states of Fe and Mn in their ground states. Furthermore, these spin states are also consistent with the size of the magnetic moments, which is larger for Mn in \ce{FeMnP_{0.55}Si_{0.45}}, both shown here and in previous studies~\cite{Hoglin2011,Miao2014}. Therefore, this study shows that the oxidation states for Fe and Mn are either 0 or 2+, with 0 being the most chemically sound oxidation state in \ce{Fe2P}-based compounds, however, the oxidation states needs other experimental validation. Similar to \ce{Fe2P} the simulated spectra for \ce{FeMnP_{0.55}Si_{0.45}} do not capture the prominent low Q feature.

In summary, the experimental INS spectra above Q~=~0.5~\AA$^{-1}$ for both \ce{Fe2P} and \ce{FeMnP_{0.55}Si_{0.45}} are well described by the linear spin wave theory simulations using the magnetic exchange parameters computed by the magnetic force theorem. Thus the simulations show that the excitations stemming from the long range ordered magnetic structure are dominated by the first six magnetic exchange parameters (J1-J6), while the lower Q feature that occurs in both systems has a different origin.

\begin{figure}[t!bh]
	\centering
		\includegraphics[width=0.99\textwidth]{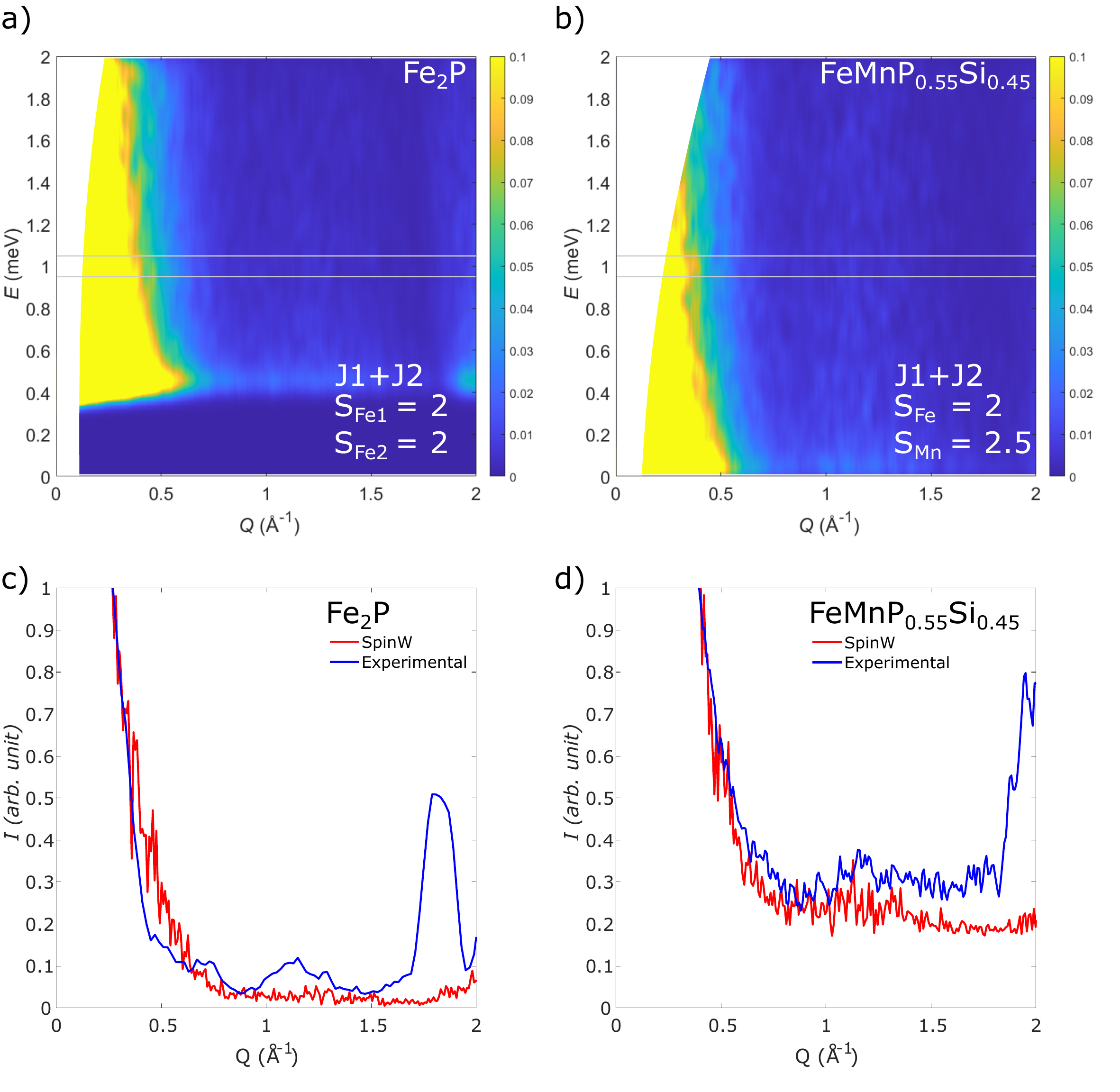}
	\caption{Simulated SpinW INS spectra using J1 and J2 for a) \ce{Fe2P} and b) \ce{FeMnP_{0.55}Si_{0.45}}. c)~Comparison between the integrated intensity (0.95 to 1.05 meV) of the experimental INS spectra at 100 K and the simulated INS spectra for \ce{Fe2P}. d) Comparison between the integrated intensity (0.95 to 1.05~meV) of the experimental INS spectra at 300~K and the simulated INS spectra for \ce{FeMnP_{0.55}Si_{0.45}}. The gray lines in a) and b) indicate the integrated regions for the simulated spectra plotted in c) and d).}
	\label{fig:LowQ}
\end{figure}

In an attempt to understand the origin of the low Q feature present in both \ce{Fe2P} and \ce{FeMnP_{0.55}Si_{0.45}} SpinW simulation which only considered a limited number of interactions where carried out. All combinations of J's that resulted in the magnetic atoms being connected in three dimensions (e.g. J1-J4 or J3+J4 in Figure~\ref{fig:Jij}) lead to a simulated spectrum with well defined correlated magnetic excitations above $\sim$0.5~\AA$^{-1}$ and a lack of a distinct low Q feature. However, simulations which considered combinations of J's that resulted in a formation of a disconnected magnetic lattice built up by isolated magnetic entities (e.g. J1+J2 or J1+J3; see Figure~\ref{fig:Jij}) resulted in prominent low Q features similar to those observed in the experimental INS spectra; see Figure \ref{fig:LowQ} a) and b). These simulations, interestingly, showed a lack of distinct magnetic excitations above $\sim$0.5~\AA$^{-1}$ as observed in the experimental spectra. No combination of J's resulted in a spectrum exhibiting both the low Q features and the distinct magnetic excitations above $\sim$0.5~\AA$^{-1}$. In order to compare the similarities of the low Q feature in the simulated spectra with the low Q feature in the experimental spectra the intensity was integrated across 0.95~\textless~$\omega$~\textless~1.05~meV and plotted as a function of Q for the simulated and experimental spectra in Figure \ref{fig:LowQ} c) and d). The simulated and experimental low Q spectra compare closely for both \ce{Fe2p} and \ce{FeMnP_{0.55}Si_{0.45}} suggesting that the origin of the low Q feature is due to the occurrence of isolated magnetic entities which mainly are dominated by the first two magnetic exchange parameters (J1 and J2).

\section{Conclusions}
In this study, we have investigated the magnetic phase transitions in the compounds \ce{Fe2P} and \ce{FeMnP_{0.55}Si_{0.45}} by using magnetometry, neutron diffraction, theoretical modelling and inelastic neutron scattering. Close to T$\rm _C$ in \ce{Fe2P}, a canted magnetic structure with magnetic moments solely on the Fe$_{3g}$-site is observed using neutron diffraction. This clearly shows that it is the Fe$_{3g}$-site that drives the magnetisation in \ce{Fe2P}, in agreement with previous theoretical calculations. In contrast, \ce{FeMnP_{0.55}Si_{0.45}} show a more gradual magnetic transition with magnetic moments on both sites. However, QENS data at low Q highlights the similar behaviour in the magnetic process in \ce{Fe2P} and \ce{FeMnP_{0.55}Si_{0.45}} with the existence of uncorrelated magnetism in both compounds below T$\rm _C$, independent of the difference in anisotropy in the compounds. This shows that anisotropy is not the most important parameter for the magnetisation process in \ce{Fe2P}-based materials.

The inelastic neutron scattering show temperature dependent features with two distinct length scales for both compounds. In the long range regime, the magnetic structure of both \ce{Fe2P} and \ce{FeMnP_{0.55}Si_{0.45}} revealed complex magnetic interactions. To fully simulate the S(\textbf{Q},$\omega$) spectra, magnetic exchange interaction parameters up to the sixth nearest neighbour (J1 to J6) needs to be considered. The $\mathbf{J_{i,j}}$'s used in the linear spin wave simulations were directly incorporated from first principles calculations, giving good agreement with the experimental S(\textbf{Q},$\omega$) spectra only needing to evaluate the spin states (S$\rm _{Fe}$ and S$\rm _{Mn}$). 

The low Q scattering could only be explained using a subset of the exchange parameters (J1 and J2), highlighting the existence of a two part system. It is the two part system that drives the magnetic transition and in turn also the magnetocaloric effect in \ce{FeMn(P{,}Si)} compounds. This emphasises that magnetic anisotropy is less important then believed previously. The existence of temperature dependant short range correlations - more distinct at T$\rm _C$ than at low temperatures - show that magnetic clusters exist upon magnetisation of the compound which upon further cooling yields the long range ordering, in agreement with Percolation theory reminiscent of spin-liquids in Kagome lattices~\cite{Li2021,Zheng2024}.

Further studies on single crystals are needed to fully resolve the findings presented here, as both the canted magnetic structure and the magnon/phonon interplay would be easier to access using single crystals. To provide a clearer understanding of the J's - neutron single crystal scattering experiments with higher energy resolution are needed to resolve possible low lying anisotropy in \ce{FeMnP_{0.55}Si_{0.45}}. Furthermore, experiments using polarised neutrons could be used to separate the contributions from phonons and magnons in the sample. As one of few experimental studies on the spin dynamics in magnetocaloric systems, this study can function as an experimental foundation for future computational and experimental studies.

\section*{Acknowledgements}
Financial support from the Swedish Research Council (VR), grant nr. 2017-06345 and 2019-00645, ÅForsk Foundation grant nr. 21-453 and 22-378 and the Swedish Foundation for Strategic Research (SSF), project ''Magnetic materials for green energy technology'' (contract EM-16-0039) are gratefully acknowledged. 

M.S.A acknowledges support from the G\"{o}ran Gustafsson Foundation.

E.K. D.-Cz. acknowledges support from the Wallenberg Initiative Materials Science for Sustainability (WISE) funded by the Knut and Alice Wallenberg Foundation (KAW), STandUPP, eSSENCE, and NL-ECO: Netherlands Initiative for Energy-Efficient Computing (with project number NWA. 1389.20.140) of the NWA research program. The computations were enabled by resources provided by the National Academic Infrastructure for Supercomputing in Sweden (NAISS), partially funded by the Swedish Research Council through grant agreement no. 2022-06725.

Experiments at the ISIS Neutron and Muon Source were supported by a beamtime allocation RB1920357 from the Science and Technology Facilities Council.
A portion of this research used resources at the Spallation Neutron Source, a DOE Office of Science User Facility operated by the Oak Ridge National Laboratory.
Furthermore, this work is based upon experiments performed at the TOF-TOF instrument operated by FRM II/ TUM at the Heinz Maier-Leibnitz Zentrum (MLZ), Garching, Germany.

\section*{CRediT author contribution statement}
\textbf{Mikael S. Andersson} Data curation, Formal analysis, Investigation, Visualisation, Writing – original draft, Writing – review \& editing
\textbf{Simon R. Larsen} Data curation, Formal analysis, Investigation, Visualisation, Writing – review \& editing
\textbf{Erna K. Delczeg-Czirjak} Data curation, Formal analysis, Investigation, Visualisation, Writing – original draft, Writing – review \& editing
\textbf{Antonio Corona} Data curation, Investigation, Writing – review \& editing
\textbf{Jacques Ollivier} Data curation, Investigation, Writing – review \& editing
\textbf{Wiebke Lohstroh} Data curation, Investigation, Writing – review \& editing
\textbf{Helen Y. Playford} Data curation, Investigation, Writing – review \& editing
\textbf{Cheng Li} Data curation, Investigation, Writing – review \& editing
\textbf{Pascale P. Deen} Conceptualisation, Data curation, Formal analysis, Investigation, Visualisation, Writing – original draft, Writing – review \& editing
\textbf{Johan Cedervall} Conceptualisation, Data curation, Formal analysis, Investigation, Project administration, Visualisation, Writing – original draft, Writing – review \& editing


\end{document}